\documentclass[12pt]{article}
\usepackage{graphicx,color}
\usepackage{XoohmE}
\usepackage{booktabs}
\def\bo{{\raise.005ex\hbox{\large$\Box$}}}

 \font\ro=cmsy10                          % font with rope
 \def\kcr{{\hbox{\ro \char'170}}}                % right-handed rope
 \def\ktl{{\hbox{\ro \char'170}}}        % top end for left-handed rope
 \def\ktr{{\hbox{\ro \char'170}}}        % " right
 \def\kbl{{\hbox{\ro \char'170}}}        % " bottom left
 \def\kbr{{\hbox{\ro \char'170}}}        % " right
 \parskip=\medskipamount                 % space between paragraphs (LaTeX)
 \thicklines                         % thick straight lines for pictures (LaTeX)

 \def\qd{{\kern0.5pt
                   q \kern-5.05pt \raise5.8pt\hbox{$\textstyle.$}\kern 0.5pt}}
\def\spin{\mathfrak{spin}}
\newdimen\parshift\parshift=\parindent
\catcode`@=11
 \long\def\@footnotetext#1{\insert\footins{\reset@font\footnotesize\interlinepenalty%
  \interfootnotelinepenalty\splittopskip\footnotesep\splitmaxdepth\dp\strutbox%
   \floatingpenalty\@MM\hsize\columnwidth\addtolength{\hsize}{-2\parindent}
    \@parboxrestore\protected@edef\@currentlabel{\csname p@footnote\endcsname\@thefnmark}
      \color@begingroup
       \@makefntext{\rule\z@\footnotesep\ignorespaces#1\@finalstrut\strutbox}
        \color@endgroup}}
 \long\def\@makefntext#1{\hglue\parshift
                         \vbox{\noindent\hb@xt@0em{\hss\@makefnmark\,}#1}}
\catcode`@=12
 \def\ba{\left(\begin{array}}
 \def\ea{\end{array}\right)}
 \def\der{\partial}
 \def\brr{\begin{eqnarray}}
 \def\err{\end{eqnarray}}

 \def\dslash{\hbox{\ooalign{$\displaystyle\partial$\cr$/$}}}
 \newcommand{\fr}[2]{{\textstyle\frac{#1}{#2}}}

 \HarvTitles %section titles more modest
 \seceq

 \def\vC#1{\vcenter{\hbox{\hss#1\hss}}}

 \def\oldheadpic{                                % old UM heading
        \setlength{\unitlength}{.4mm}
        \thinlines
        \par
        \begin{picture}(349,16)
        \put(325,16){\line(1,0){4}}
        \put(330,16){\line(1,0){4}}
        \put(340,16){\line(1,0){4}}
        \put(335,0){\line(1,0){4}}
        \put(340,0){\line(1,0){4}}
        \put(345,0){\line(1,0){4}}
        \put(329,0){\line(0,1){16}}
        \put(330,0){\line(0,1){16}}
        \put(339,0){\line(0,1){16}}
        \put(340,0){\line(0,1){16}}
        \put(344,0){\line(0,1){16}}
        \put(345,0){\line(0,1){16}}
        \put(329,16){\oval(8,32)[bl]}
        \put(330,16){\oval(8,32)[br]}
        \put(339,0){\oval(8,32)[tl]}
        \put(345,0){\oval(8,32)[tr]}
        \end{picture}
        \par
        \thicklines
        \vskip.2in}
 \def\oldtitle#1#2#3#4{\oldheadpic\begin{center}\vglue.5in{\large\bf #1}\\[.6in]
        {#2}\\[.1in] {\it Department of Physics and Astronomy}\\
        {\it University of Maryland, College Park, MD 20742}\\[.6in]
        Physics Publication \#{#3}\\ {#4}\\[1.5in] {\bf ABSTRACT}\\[.1in]
        \end{center} \begin{quotation}}                 % old title stuff
 \def\oldTitle#1#2#3#4#5#6#7{\oldheadpic\begin{center} \vglue .4in
        {\large\bf #1}\\[.4in]
        {#2}\\[.1in] {\it Department of Physics and Astronomy}\\
        {\it University of Maryland, College Park, MD 20742}\\[.1in]
        {#3}\\[.1in] {\it {#4}}\\ {\it {#5}}\\[.4in]
        Physics Publication \#{#6}\\ {#7}\\[.5in] {\bf ABSTRACT}\\[.1in]
        \end{center} \begin{quotation}}                 % " for 2 authors
 \def\border{                                            % border
        \setlength{\unitlength}{1mm}
        \newcount\xco
        \newcount\yco
        \xco=-21
        \yco=12
        \begin{picture}(140,0)
        \put(\xco,\yco){$\ktl$}
        \advance\yco by-1
        {\loop
        \put(\xco,\yco){$\kcr$}
        \advance\yco by-2
        \ifnum\yco>-240
        \repeat
        \put(\xco,\yco){$\kbl$}}
        \xco=158
        \yco=12
        \put(\xco,\yco){$\ktr$}
        \advance\yco by-1
        {\loop
        \put(\xco,\yco){$\kcr$}
        \advance\yco by-2
        \ifnum\yco>-240
        \repeat
        \put(\xco,\yco){$\kbr$}}
        \put(-20,13){\tiny ***
        SUNY Oneonta ** Physics Department **
        SUNY Oneonta *** Physics Department ***
        SUNY Oneonta *** Physics Department ***
        SUNY Oneonta **}
        \put(-20,-241.5){\tiny **University of Maryland * Center for String and
        Particle  Theory* Physics Department***University of Maryland*Center
        for String and Particle Theory** }
        \end{picture}
        \par\vskip-8mm}
 \def\bordero{                                           % alternate border
        \setlength{\unitlength}{1mm}
        \newcount\xco
        \newcount\yco
        \xco=-31
        \yco=12
        \begin{picture}(140,0)
        \put(\xco,\yco){$\ktl$}
        \advance\yco by-1
        {\loop
        \put(\xco,\yco){$\kclr}
        \advance\yco by-2
        \ifnum\yco>-240
        \repeat
        \put(\xco,\yco){$\kbl$}}
        \xco=151
        \yco=12
        \put(\xco,\yco){$\ktr$}
        \advance\yco by-1
        {\loop
        \put(\xco,\yco){$\kcr$}
        \advance\yco by-2
        \ifnum\yco>-240
        \repeat
        \put(\xco,\yco){$\kbr$}}
        \put(-20,12){\ooo bacdefghidfghghdhededbihdgdfdfhhdheidhdhebaaahjhhdahbahgdedgehgfdiehhgdigicba}
        \put(-20,-241.5){\oooababaighefdbfghgeahgdfgafagihdidihiidhiagfedhadbfdecdcdfagdcbhaddhbgfchbgfdacfediacbabab}
        \end{picture}
        \par\vskip-8mm}
 \def\headpic{                                           % UM heading
        \indent
        \setlength{\unitlength}{.4mm}
        \thinlines
        \par
        \begin{picture}(29,16)
        \put(165,16){\line(1,0){4}}
        \put(170,16){\line(1,0){4}}
        \put(180,16){\line(1,0){4}}
        \put(175,0){\line(1,0){4}}
        \put(180,0){\line(1,0){4}}
        \put(185,0){\line(1,0){4}}
        \put(169,0){\line(0,1){16}}
        \put(170,0){\line(0,1){16}}
        \put(179,0){\line(0,1){16}}
        \put(180,0){\line(0,1){16}}
        \put(184,0){\line(0,1){16}}
        \put(185,0){\line(0,1){16}}
        \put(169,16){\oval(8,32)[bl]}
        \put(170,16){\oval(8,32)[br]}
        \put(179,0){\oval(8,32)[tl]}
        \put(185,0){\oval(8,32)[tr]}
        \end{picture}
        \par\vskip-6.5mm
        \thicklines}
 \def\title#1#2#3#4{\border\headpic {\hbox to\hsize{#4 \hfill UMDEPP #3}}\par
        \begin{center} \vglue .5in {\large\bf #1}\\[.6in]
        {#2}\\[.1in] {\it Department of Physics and Astronomy}\\
        {\it University of Maryland, College Park, MD 20742}\\[1.5in]
        {\bf ABSTRACT}\\[.1in] \end{center} \begin{quotation}}  % title stuff
 \def\Title#1#2#3#4#5#6#7{\border\headpic
        {\hbox to\hsize{#7 \hfill UMDEPP #6}}\par
        \begin{center} \vglue .4in {\large\bf #1}\\[.4in]
        {#2}\\[.1in] {\it Department of Physics and Astronomy}\\
        {\it University of Maryland, College Park, MD 20742}\\[.1in]
        {#3}\\[.1in] {\it {#4}}\\ {\it {#5}}\\[.5in] {\bf ABSTRACT}\\[.1in]
        \end{center} \begin{quotation}}                 % " for 2 authors
 \def\endtitle{\end{quotation}\newpage}                  % end title page

 \def\qd{{\kern0.5pt
                   q \kern-5.05pt \raise5.8pt\hbox{$\textstyle.$}\kern 0.5pt}}

\begin{document}
 \border\headpic {September 2008}
 {\hfill {UMDEPP-08-018}}\\[-.5in]
 {\flushright{\hfill {SUNY-O/667}}}

 \vspace{.4in}

 \setlength{\oddsidemargin}{0.3in}
 \setlength{\evensidemargin}{-0.3in}
 \begin{center}
 \vglue .05in {\Large\bf Frames for Supersymmetry }
 \\[.25in]

   { C.F.\,Doran$^a$, M.G.\,Faux$^b$, S.J.\,Gates, Jr.$^c$,
     T.\,H\"{u}bsch$^d$, K.M.\,Iga$^e$, G.D.\,Landweber$^f$}\\[3mm]
{\small\it
  $^a$Department of Mathematical and Statistical Sciences, University of Alberta,
  Edmonton, Alberta T6G 2G1 CANADA\\[-1mm]
  {\tt  doran@math.washington.edu}
  \\
  $^b$Department of Physics,
      State University of New York, Oneonta, NY 13820\\[-1mm]
  {\tt  fauxmg@oneonta.edu}
  \\
  $^c$Department of Physics,
      University of Maryland, College Park, MD 20472\\[-1mm]
  {\tt  gatess@wam.umd.edu}
  \\
  $^d$Department of Physics and Astronomy,
      Howard University, Washington, DC 20059\\[-1mm]
  {\tt  thubsch@mac.com}
  \\
  $^e$Natural Science Division,
      Pepperdine University, Malibu, CA 90263\\[-1mm]
  {\tt  Kevin.Iga@pepperdine.edu}
  \\
 $^f$Department of Mathematics, Bard College, Annandale-on-Hudson, NY
 12504-5000\\[-1mm]
  {\tt gregland@bard.edu}
 }\\[9mm]

 {\bf ABSTRACT}\\[.01in]
 \end{center}
 \begin{quotation}
 {We explain how the redefinitions of supermultiplet component fields,
 comprising what we call ``frame shifts", can be used in conjuction with
 the graphical technology of multiplet Adkinras to render manifest
 the reducibility of off-shell representations of supersymmetry.  This technology
 speaks to possibility of organizing multiplet constraints in
 a way which complements and extends the possibilities afforded by
 superspace methods.}

 ${~~~}$ \newline PACS: 04.65.+e

 \endtitle
  Four years ago, in \cite{FG1}, graphical devices for representing
 supermultiplets were introduced.  We have used these tools frequently
 since that time, as we find these distinctly useful for
 organizing various open questions about supersymmetry.
 Moreover, these graphs illuminate intriguing emergent mathematical features of
 supersymmetry not manifest in the context of more traditional methods, such as
 superspace.  Accordingly, we
 have been incrementally adding mathematical sophistication to this technology
 \cite{Prepotentials,DFGHIL01,AT1,HDS}.
 For reasons explained in \cite{FG1}, we refer to our graphical representations
 of supermultiplets as Adinkras.  In the case of one-dimensional
 supersymmetry we have developed a means, reminiscent of Feynman
 rules, which allows unambiguous translation of supersymmetry
 transformations from these graphs.  In four-dimensions the
 technology is less-developed, and the graphs serve more as
 helpful visual aids.  However, there is an especially useful feature
 exhibited by four-dimensional Adinkras concerning how
 the graphical structures may
 be manipulated to expose and render obvious when and how the multiplet
 admits projection to submultiplets.  The primary goal of this letter
 is to explain this.

 We should point out that the analysis of the reduction of $N=1$ supermultiplets
 described in this letter is an old and well known story, described in many places,
 e.g. \cite{Superspace,Wess_Bagger}.  But the methodology
 which we bring to bear on this problem is new and interesting.
 We believe that practitioners of supersymmetry will appreciate the fresh look
 that this perspective brings to this matter, especially as regards its potential for
 resolving related issues in higher-$N$ supersymmetry, where superspace methods become
 increasingly cumbersome.

 The basic idea behind Adinkras is a deceptively simple one:
 to construct graphs by representing fields as vertices which are
 interconnected pairwise by edges when the corresponding fields are
 related by supersymmetry transformations.  No-doubt practitioners have
 done this sort of thing ever since supersymmetry was first
 conceived in the early 1970s.  However, it has become increasingly
 evident that such diagrams encode a wealth of
 information which may be extracted beneficially to complement or even
 replace some traditional methods for organizing and
 classifying supermultiplets and supersymmetric actions.
 As a notable example, the one-dimensional Adinkras obtained
 by dimensional reduction of supermultiplets in any number of
 dimensions admit topological classification in terms of
  doubly-even linear binary codes \cite{HDS}.

 One of our prime motivations has been to better
 understand the nexus of ways in which higher-$N$
 theories may be realized off-shell.  One preliminary result
 obtained using our methods was the discovery of a way
 to couple what we call quadruplet matter to $N=2$
 hypermultiplets, using a finite number of off-shell
 degrees of freedom.  This was explained in \cite{RETM}.
 That work exposed deeper questions, which we hope to resolve, concerning the
 possibility of removing the quadruplet matter to expose an interesting
 new off-shell realization of a pure hypermultiplet.  Our scrutiny of this question impelled
 us investigate structures in \cite{RETM} using the simpler setting of $N=1$ supersymmetry;
 the results of this letter derived from that investigation.
  All of this has a deeper
 motivation related to our desire to develop either an
 off-shell realization of $N=4$ Super Yang-Mills theory involving a finite number
 of component fields, or to prove that such constructions are
 precluded within the ordinary framework of supersymmetric quantum field
 theory.  The latter is a commonly-held belief which we have not found
 demonstrably substantiated in the literature.

 As explained above, an Adinkra consists of a set of vertices, one for
 each component field in a given supermultiplet.  These vertices are connected pairwise by
 edges when the
 corresponding component fields are connected by a supersymmetry transformation.
 We color boson vertices white, and we color fermion
 vertices black.
 In four-dimensions, multiplet component fields comprise irreducible representations of
 $\spin(3,1)$; we use a single vertex for each such field and
 decorate this with a numeral to indicate the number of off-shell degrees of
 freedom described by that field.
 For example, a complex scalar boson would be represented
 by a white vertex decorated with the numeral 2.  Next, we organize the vertices
 vertically in a manner which faithfully respects the engineering dimension of the
 fields, with lower-dimension fields at the bottom of the diagram and higher-dimension fields
 placed at successively higher ``levels".

 As a simple example, the Chiral multiplet consists of a complex scalar
 $\phi$, a right-handed Weyl spinor $\psi_R$, and a higher-weight complex scalar $F$.
 This multiplet has the following supersymmetry transformation rules,
  \brr \delta_Q\,\phi &=&
      i\,\bar{\e}_L\,\psi_R
      \nonumber\\[.1in]
      \delta_Q\,\psi_R &=&
      \dslash\,\phi\,\e_L
      +F\,\e_R
      \nonumber\\[.1in]
      \delta_Q\,F &=&
      i\,\bar{\e}_R\,\dslash\,\psi_R \,,
 \label{cc}\err
 where $\e_L$ is a left-handed Weyl spinor supersymmetry parameter and
 $\e_R$ is its Majorana conjugate.
 We represent this multiplet by the following Adinkra,
 \begin{equation}
   \vC{\includegraphics[width=.7in]{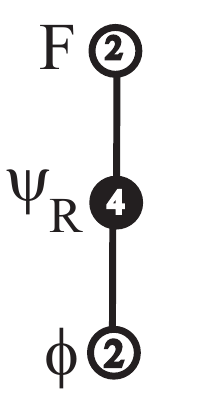}} \,.
 \label{cca}\end{equation}
 Under (\ref{cc}) the complex scalar $\phi$ transforms into the spinor $\psi_R$,
 and the spinor $\psi_R$ transforms into the derivative of the
 scalar; for these reasons the $\phi$ vertex and the
 $\psi_R$ vertex are connected by a black edge in the diagram (\ref{cca}).  Similarly, the spinor
 $\psi_R$ transforms into the complex scalar $F$ while this field
 transforms into the derivative of the spinor; for these
 reasons the $\psi_R$ node is connected by a black edge to the
 $F$ node.

 In the Chiral Adinkra (\ref{cca}) each edge represents two terms in the
 transformation rules (\ref{cc}): one ``upward-directed" under which a lower-weight field
 transforms into a higher-weight field, and one ``downward directed"
 in which a higher-weight field transforms into
 the derivative of the lower-weight field. Conventionally,
 the black-coloration of an edge denotes this bidirectional
 quality.  As it turns out, it is
 possible to structure some transformation rules so that
 a given upward-directed term does not have a downward-directed
 counterpart; we give explicit examples of this below.  In such a
 circumstance, we denote an upward-directed term which does not have a downward-directed
 counterpart using a grey, rather than a black, edge.

 Consider next a Real Scalar multiplet $V$
 as described by the following component supersymmetry transformation rules,
 \brr \d_Q\,b &=& \fr12\,i\,\bar{\e}\,\chi
      \nonumber\\[.1in]
      \d_Q\,\chi &=& \dslash\,b\,\e
      +\fr12\,i\,A_{a}\,\gamma^a\g_5\,\e
      +\fr12\,f\,\e
      -\fr12\,i\,g\,\g_5\,\e
      \nonumber\\[.1in]
      \d_Q\,f &=& \fr12\,i\,\bar{\e}\,\dslash\chi
      +\fr12\,i\,\bar{\e}\,\lambda
      \nonumber\\[.1in]
      \d_Q\,g &=& \fr12\,\bar{\e}\,\g_5\,\dslash\chi
      +\fr12\,\bar{\e}\,\g_5\lambda
      \nonumber\\[.1in]
      \d_Q\,A_a &=&
      \fr12\,\bar{\e}\,\g_5\,\dslash\g_a\,\chi
      -\fr12\,\bar{\e}\,\g_a\g_5\,\lambda
      \nonumber\\[.1in]
      \d_Q\lambda &=&
      \fr12\,\dslash f\,\e
      -\fr12\,i\,\dslash g\,\g_5\,\e
      +\fr12\,i\,\gamma^a\dslash\,A_a\,\g_5\,\e
      +D\,\e
      \nonumber\\[.1in]
      \d_Q\,D &=&
      \fr12\,i\,\bar{\e}\,\dslash\,\lambda \,.
 \label{rsm1}\err
 where the scalar component field $b$ is assigned even parity.  The parity of all other components
 then follows from the requirement that a parity flip commute with supersymmetry; for example
 $\chi$ and $\lambda$ each have even parity, $g$ is a pseudoscalar, and $A_a$ is an axial vector.

 Consider also a Real Pseudoscalar multiplet $\tilde{V}$
 as described by the following component supersymmetry transformation rules,
 \brr \d_Q\,\tilde{b} &=& \fr12\,\bar{\e}\,\gamma_5\,\tilde{\chi}
      \nonumber\\[.1in]
      \d_Q\,\tilde{\chi} &=& i\,\gamma_5\dslash\,\tilde{b}\,\e
      -\fr12\,V_a\,\gamma^a\,\e
      +\fr12\,\tilde{f}\,\e
      -\fr12\,i\,\tilde{g}\,\g_5\,\e
      \nonumber\\[.1in]
      \d_Q\,\tilde{f} &=& \fr12\,i\,\bar{\e}\,\dslash\tilde{\chi}
      +\fr12\,i\,\bar{\e}\,\tilde{\lambda}
      \nonumber\\[.1in]
      \d_Q\,\tilde{g} &=& \fr12\,\bar{\e}\,\g_5\,\dslash\tilde{\chi}
      +\fr12\,\bar{\e}\,\g_5\tilde{\lambda}
      \nonumber\\[.1in]
      \d_Q\,V_a &=&
      \fr12\,i\,\bar{\e}\,\dslash\g_a\,\tilde{\chi}
      +\fr12\,i\,\bar{\e}\,\g_a\,\tilde{\lambda}
      \nonumber\\[.1in]
      \d_Q\,\tilde{\lambda} &=&
      \fr12\,\dslash \tilde{f}\,\e
      +\fr12\,i\,\gamma_5\dslash \tilde{g}\,\e
      -\fr12\,\gamma^a\dslash\,V_a\,\e
      -i\,\tilde{D}\,\gamma_5\,\e
      \nonumber\\[.1in]
      \d_Q\,\tilde{D} &=&
      \fr12\,\bar{\e}\,\gamma_5\dslash\,\tilde{\lambda} \,.
 \label{rpsm1}\err
 The distinction between this multiplet and (\ref{rsm1}) pertains to the parity of
 the lowest component scalar fields; in this case the component $\tilde{b}$ is assigned odd
 parity.  It follows that $\tilde{\chi}$ and $\tilde{\lambda}$ each have even parity, the fields $\tilde{g}$ and $\tilde{D}$ are each pseudoscalars,
 while the vector $V_a$ has even parity.\footnote{As far as supersymmetry is concerned, the Real Scalar
 multiplet (\ref{rsm1}) and the Real
 Pseudoscalar multiplet (\ref{rpsm1}) describe the same representation.  This can be readily shown
 by redefining the spinor components in the former by a cosmetic multiplication by $i\,\gamma_5$
 and by making other minor cosmetic changes.  The conventional difference ensures that all Majorana
 spinors in both multiplets have even parity.}

 Taken together, the transformation rules (\ref{rsm1}) and (\ref{rpsm1}) are represented
 diagrammatically as in Figure \ref{clma},
 \begin{figure}
 \begin{center}
 \includegraphics[width=3.5in]{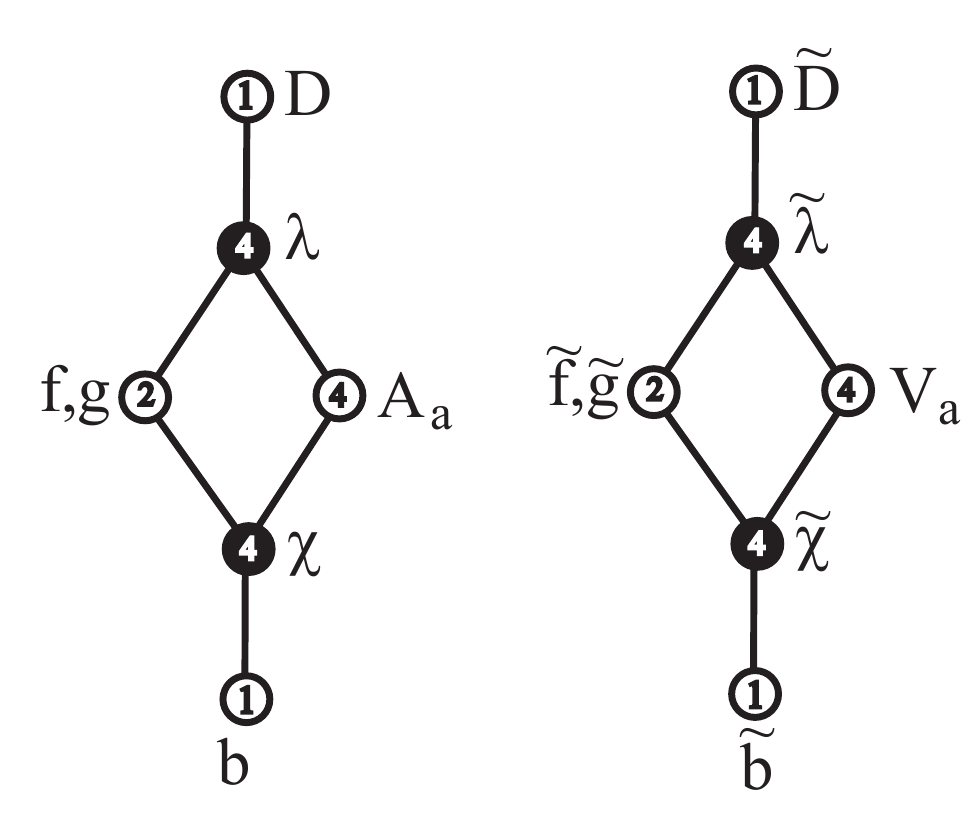}
 \caption{The real and imaginary parts of a Complex Scalar multiplet
 correspond to a Real Scalar multiplet (\ref{rsm1}) and a Real Pseudoscalar multiplet
 (\ref{rpsm1}), respectively.  These are represented above using a disconnected Adinkra.}
 \label{clma}
 \end{center}
 \end{figure}
 where a single combined vertex
 represents $f$ and $g$ together, and another combined
 vertex represents $\tilde{f}$ and $\tilde{g}$.
 The fact that Figure \ref{clma} has two disconnected parts reflects the
 fact that the fields in the multiplet $V$ do not transform into
 the fields in $\tilde{V}$, and vice-versa.
 This pair admits an obvious complex structure allowing us to
 define ${\mathbb V}:=V+i\,\tilde{V}$.  The reducibility of the combined
 mutliplet is manifest using the diagram in Figure \ref{clma}, since one
 can constrain all of the fields in either disconnected
 piece to vanish, an operation which is consistent with
 supersymmetry owing to the disconnected feature of the diagram.

 The transformation rules (\ref{rsm1}) and (\ref{rpsm1}) describe together the
 component version of the transformations generated by the supercharge
 $Q$ acting on an unconstrained $N=1$ superfield corresponding to
 ${\mathbb V}$. From the point of view of superspace, the restriction to the
 submultiplet described by the connected diagram on the left-side of
 Figure \ref{clma}, effected by constraining the connected diagram on the right-side to
 vanish, is equivalent to the superfield constraint ${\mathbb V}={\mathbb V}^\dagger$, where the
 complex structure described above has been implied.  Similarly, the
 restriction to the other connected sub-multiplet
 is equivalent to the superspace constraint ${\mathbb V}= -{\mathbb V}^\dagger$.

 Since we have assigned positive parity to $V$ and
 negative parity to $\tilde{V}$, so that $b$ and $\tilde{b}$ describe
 a respective scalar and a pseudoscalar, it follows that we can sensibly
 reorganize the component fields using the following definitions,
 \brr B &:=&
      b-i\,\tilde{b}
      \nonumber\\[.1in]
      \xi_L &:=&
      \fr12\,(\,\chi_{L}+\tilde{\chi}_L\,)
      \nonumber\\[.1in]
      \rho_R &:=&
      \fr12\,(\,\chi_R-\tilde{\chi}_R\,)
      \nonumber\\[.1in]
      P_a &:=&
      -\fr14\,(\,V_a-i\,A_a\,)
      +\fr12\,\der_a\,(\,b-i\,\tilde{b}\,)
      \nonumber\\[.1in]
      H &:=&
      -\fr12\,(\,\tilde{f}-i\,\tilde{g}\,)
      \nonumber\\[.1in]
      \beta_L &:=&
      \fr14\,(\,\lambda_L-\dslash\,\chi_R\,)
      +\fr14\,(\,\tilde{\lambda}-\dslash\,\tilde{\chi}_R\,)
      \nonumber\\[.1in]
      \phi &:=&
      \fr12\,(\,f+\tilde{f}\,)
      +\fr12\,i\,(\,g+\tilde{g}\,)
      \nonumber\\[.1in]
      \psi_R &:=&
      \fr12\,(\,\lambda_R+\dslash\,\chi_L\,)
      +\fr12\,(\,\tilde{\lambda}_R+\dslash\,\tilde{\chi}_L\,)
      \nonumber\\[.1in]
      F &:=& \fr12\,(\,D+\der^aV_a-\Box\,b\,)
      -\fr12\,i\,(\,\tilde{D}+\der^aA_a-\Box\,\tilde{b}\,) \,.
 \label{Chi_In}\err
 By replacing the fields
 $(\,b\,,\,\chi\,,\,f\,,\,g\,,\,A_a\,,\,\lambda\,,\,D\,)$
 and $(\,\tilde{b}\,,\,\tilde{\chi}\,,\,\tilde{f}\,,\,\tilde{g}\,,\,V_a\,,\,\tilde{\lambda}\,,\,\tilde{D}\,)$
 with the equivalent set of fields defined by
 $(\,B\,,\,\rho\,,\,H\,,\,\xi\,,\,P_a\,,\,\beta\,,\,\phi\,,\,\psi\,,\,F\,)$
 we have ``changed frames" in the space defined by these
 fields; in either guise the same 16+16 local off-shell degrees of freedom
 are expressed, albeit in terms of different linear combinations.\footnote{If a parity operation should act canonically,
 as $P:\,\xi_L\leftrightarrow \xi_R$, then the parity of
 $V$ and $\tilde{V}$ must be opposite.  Thus, if
 $b$ is a scalar then $\tilde{b}$ must be a pseudoscalar.  This justifies the assignments
 imposed above.}
 When re-expressed in terms of the redefined fields, the transformation rules (\ref{rsm1}) become
 \brr \delta_Q\,B &=&
      i\,\bar{\e}_L\,\rho_R
      +i\,\bar{\e}_R\,\xi_L
      \nonumber\\[.1in]
      \delta_Q\,\xi_L &=&
      P_a\,\gamma^a\,\e_R
      +\fr12\,\phi\,\e_L
      \nonumber\\[.1in]
      \delta_Q\,\rho_R &=&
      \dslash\,B\,\e_L
      -P_a\,\gamma^a\,\e_L
      +H\,\e_R
      \nonumber\\[.1in]
      \delta_Q\,P_a &=&
      -\fr12\,i\,\bar{\e}_R\,\dslash\gamma_a\,\xi_L
      +\fr12\,i\,\bar{\e}_L\,\gamma_a\,\beta_L
      -\fr14\,i\,\bar{\e}\,\gamma_a\,\psi
      \nonumber\\[.1in]
      \delta_Q\,H &=&
      i\,\bar{\e}_R\,\dslash\,\rho_R
      +i\,\bar{\e}_R\,\beta_L
      \nonumber\\[.1in]
      \delta_Q\,\beta_L &=&
      -\gamma^a\dslash\,P_a\,\e_L
      +\fr12\,F\,\e_L
      \nonumber\\[.1in]
      \delta_Q\,\phi &=&
      i\,\bar{\e}_L\,\psi_R
      \nonumber\\[.1in]
      \delta_Q\,\psi_R &=&
      \dslash\,\phi\,\e_L
      +F\,\e_R
      \nonumber\\[.1in]
      \delta_Q\,F &=&
      i\,\bar{\e}_R\,\dslash\,\psi_R \,.
 \label{new_rules}\err
 Note that the transformation rules (\ref{rsm1}) and
 (\ref{new_rules}) are completely equivalent; these represent the same
 multiplet in two different ``frames".  The second guise for the
 transformation rules, (\ref{new_rules}), is represented
 diagrammatically as in Figure \ref{k2}.
 \begin{figure}
 \begin{center}
 \includegraphics[width=2.2in]{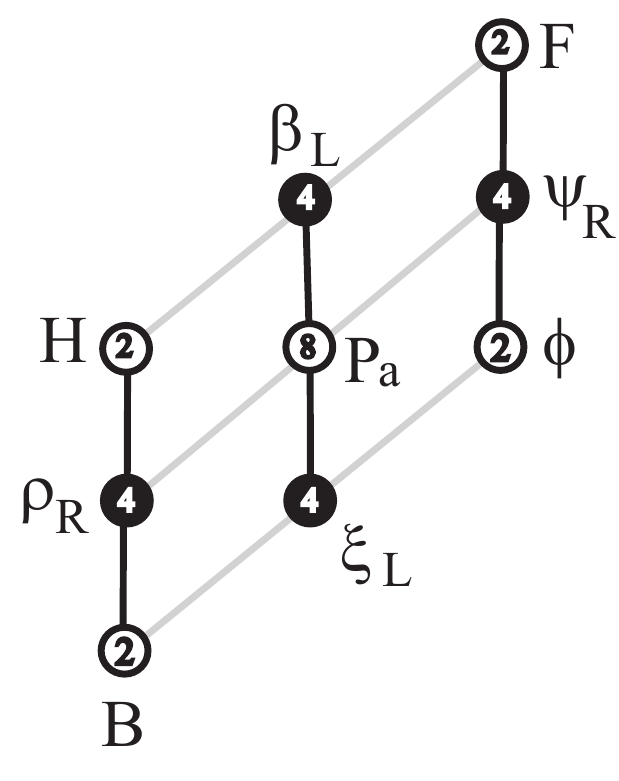}
 \caption{The Complex Scalar multiplet may be expressed in a special ``frame" in
 which the Adinkra has the manifest structure of a Chiral multiplet
 linked to a Variant Vector multiplet in-turn linked to a higher-weight
 Chiral multiplet, where the linking are codified by ``one-way" grey edges,
 as explained in the text.}
 \label{k2}
 \end{center}
 \end{figure}
 In Figure \ref{k2}, the grey edges describe upward-directed terms in
 (\ref{new_rules}) which do not have downward-directed counterparts.
 For example the grey edge connecting the $B$ vertex to the
 $\xi_L$ vertex represents the term $i\,\bar{\e}_R\,\xi_L$ appearing
 in $\delta_Q\,B$.  The fact that this edge is grey, rather than black,
 indicates the interesting fact that there is no term proportional to $B$ which
 appears in $\delta_Q\,\xi_L$; thus, the upward-directed term does not have a
 downward-directed counterpart.  Similar comments apply to all six of the grey edges
 in \ref{k2}.

 The ostensibly distinct Adinkras in Figures \ref{clma} and \ref{k2} describe precisely the same multiplet
 expressed in two different frames; these frames are related by the field redefinition
 (\ref{Chi_In}). In the second frame, the graph shown in Figure \ref{k2} exhibits
 an interesting structure:  The fields $B$, $\rho_R$, and $H$ transform into each other
 via two bi-directional term pairs represented by the two black edges which interconnect
 the corresponding vertices, but none of the remaining fields transform into
 $B$, $\rho_R$, or $H$.  Instead, the three upward-directed terms in (\ref{new_rules}) represented
 by the three grey edges connecting $B$ with $\xi_L$, $\rho_R$ with $P_a$, and $H$ with $\beta_L$,
 have no downward-directed counterparts.  Thus, the
 subset of fields $(\,B\,,\,\rho_R\,,\,H\,)$ connects to $(\,\xi_L\,,\,P_a\,,\,\beta_L\,)$
 only via grey edges.  In a similar way, the subset $(\,\xi_L\,,\,P_a\,,\,\beta_L\,)$
 connects to $(\,\phi\,,\,\psi_R\,,\,F\,)$ only by grey edges.

 The subset of fields $(\,B\,,\,\rho_R\,,\,H\,)$ have transformation rules identical
 to that of a Chiral multiplet augmented by the addition of three ``one-way" terms
 corresponding to grey edges.  As a suggestive mnemonic, we refer to this situation
 by calling $(\,B\,,\,\rho_R\,,\,H\,)$ a Chiral multiplet ``flying" the fields
 $(\,\xi_L\,,\,P_a\,,\,\beta_L\,)$, as a kite.  Similarly, the subset of fields
 $(\,\xi_L\,,\,P_a\,,\,\beta_L\,)$ have transformation rules identical to a Variant Vector multiplet
 \cite{Variant} augmented by the addition of three ``one-way" terms corresponding to the remaining
 three grey edges.\footnote{The Variant Vector multiplet may be viewed
 as a pair of Chiral multiplets, with swapped statistics, spanning a Weyl spinor
 representation of $\spin(3,1)$.}
 Thus, from this point of view, when expressed in this frame, the Complex Scalar multiplet
 is a Chiral multiplet ``flying" a Variant Vector multiplet which, in turn, is ``flying"
 another Chiral multiplet.

 This structuring of the multiplet renders manifest the following constraint which restricts (\ref{new_rules})
 to a proper submultiplet: the fields $(\,\phi\,,\,\psi_R\,,\,F\,)$ can be eliminated
 by constraining $\phi=0$, $\psi=0$, and $F=0$.  Since none of these three fields transform into any
 of the other fields in figure \ref{k2}, as clearly indicated by the grey lines, this constraint is
 plainly consistent with
 supersymmetry.  In this way, the structure of Figure \ref{k2} allows us to ``read-off" the
 Adinkra describing the constrained multiplet as
 \begin{equation}
   \vC{\includegraphics[width=1.5in]{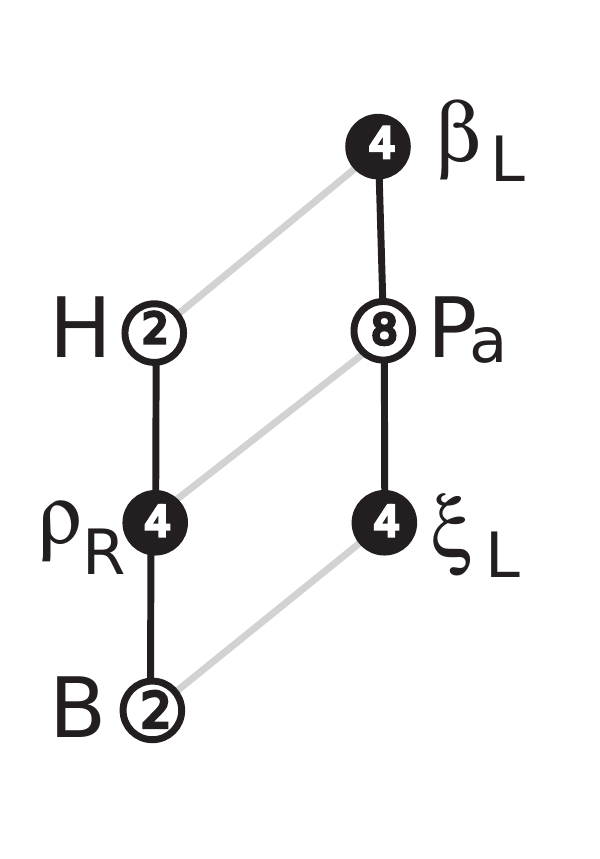}} \,.
 \label{clmsub}\end{equation}
 To pass from Figure \ref{k2} to (\ref{clmsub}) we have ``switched off" the ``uppermost kite",
 by constraining $\phi$, $\psi_R$, and $F$, each to vanish.
 The transformation rules corresponding to (\ref{clmsub}) describe a
 Complex Linear multiplet \cite{Nonminimal}, obtained equivalently in superspace by imposing
 the constraint $\bar{D}_R\,D_L\,{\mathbb V}=0$ and then redefining component fields
 according to the frame shift indicated by (\ref{Chi_In}).

 The structure of the Complex Linear Adinkra (\ref{clmsub}) renders manifest a second constraint
 that can further reduce the system to a smaller proper submultiplet: the
 fields $(\,\xi_L\,,\,P_a\,,\,\beta_L\,)$ can be eliminated by constraining
 $\xi_L=0$, $P_a=0$, and $\beta_L=0$.  This second constraint is also
 manifestly consistent with supersymmetry as indicated by the grey lines,
 since the newly constrained fields do not transform into any of the unconstrained fields.
 It is then easy to ``read off" the Adinkra describing the further
 constrained multiplet as
 \brr \vC{\includegraphics[width=0.7in]{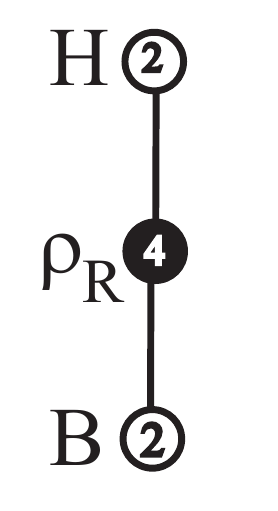}} \,.
 \label{c2ad}\err
 To pass from (\ref{clmsub}) to (\ref{c2ad}) we have ``switched off" the ``middle kite",
 by constraining $\xi_L$, $P_a$, and $\beta_L$, each to vanish.
 The transformation rules corresponding to (\ref{clmsub}) describe a
 Chiral multiplet, as could also be inferred since (\ref{c2ad})
 is similar to (\ref{cc}).
 It is simple to check that the transformation rules in (\ref{new_rules})
 for $B$, $\rho_R$, and $H$ correspond to a Chiral multiplet when
 all of the other fields are constrained to vanish.  This same Chiral multiplet is obtained
 equivalently in superspace by imposing the very well-known constraint
 $D_L\,{\mathbb V}=0$ and then redefining component fields according to the frame shift indicated by
 (\ref{Chi_In}).

 From the point of view of superspace, if one starts with an unconstrained complex superfield ${\mathbb V}$,
 this can be reduced to a 12+12 Complex Linear multiplet by imposing $\bar{D}_R\,D_L\,{\mathbb V}=0$, and
 can be further constrained to a Chiral multiplet by imposing the more restrictive constraint $D_L\,{\mathbb V}=0$.
 These constraints may be resolved in terms of components, theta-level by theta-level in
 a superfield component expansion.  Alternatively, these may be imposed in the basis defined by (\ref{Chi_In}), in which case
 the restriction corresponds to the
 diagrams in Figure \ref{k2}, in (\ref{clmsub}), and in (\ref{c2ad}).  Thus, the essence of our discussion pertains
 to the elucidation of natural frames for discussing multiplet reduction in terms of components,
 how different frames can be used for reduction to different submultiplets, and how the naturalness
 of the frames are clarified by rendering the transformation rules diagrammatically.

 In the context of $N=1$ supersymmetry, the 16+16 Complex Scalar multiplet, described above, provides
 the simplest example of multiplet reducibility, and provides an archetype for other examples.
 It is well-known that the 8+8 Real Scalar multiplet, which corresponds to either of the disconnected
 sub-Adinkras in Figure \ref{clma}, is also reducible; for example, the 4+4 Gauge Vector multiplet may be obtained by
 a suitable restriction of the component fields in this case. However, this reduction, which coincides
 with the well-known restriction to a Wess-Zumino gauge, is more subtle than either the reduction from the
 Complex Scalar multiplet to a Complex Linear multiplet or the further reduction to a Chiral multiplet.
 The reason for the extra subtlety has to do with the presence of a gauge equivalence
 associated with the Vector multiplet. To see this, consider the Real Pseudoscalar multiplet,
 obtained from equation (\ref{rsm1}) or Figure \ref{clma} by setting all fields in $V$ to vanish.
 In this way we restrict to the disconnected diagram comprising the right half
 of Figure \ref{clma}.  A natural frame
 for further reduction is then obtained by redefining fields as
 $\tilde{\lambda}\to\tilde{\lambda}+\dslash\,\tilde{\chi}$ and $\tilde{D}\to \tilde{D}-\Box\,\tilde{b}$.
 In terms of the redefined fields, the
 transformation rules  become
  \brr \d_Q\,\tilde{b} &=& \fr12\,\bar{\e}\,\gamma_5\,\tilde{\chi}
      \nonumber\\[.1in]
      \d_Q\,\tilde{\chi} &=& i\,\gamma_5\dslash\,\tilde{b}\,\e
      -\fr12\,V_a\,\gamma^a\,\e
      +\fr12\,\tilde{f}\,\e
      -\fr12\,i\,\tilde{g}\,\g_5\,\e
      \nonumber\\[.1in]
      \d_Q\,\tilde{f} &=& i\,\bar{\e}\,\dslash\tilde{\chi}
      +\fr12\,i\,\bar{\e}\,\tilde{\lambda}
      \nonumber\\[.1in]
      \d_Q\,\tilde{g} &=& \bar{\e}\,\g_5\,\dslash\tilde{\chi}
      +\fr12\,\bar{\e}\,\g_5\tilde{\lambda}
      \nonumber\\[.1in]
      \d_Q\,V_a &=&
      \fr12\,i\,\bar{\e}\,\gamma_a\,\tilde{\lambda}
      +\der_a\,(-i\,\bar{\e}\,\tilde{\chi}\,)
      \nonumber\\[.1in]
      \d_Q\,\tilde{\lambda} &=&
      \fr12\,\gamma^{ab}\,F_{ab}\,\e
      -i\,\tilde{D}\,\gamma_5\,\e
      \nonumber\\[.1in]
      \d_Q\,\tilde{D} &=&
      \fr12\,\bar{\e}\,\gamma_5\dslash\,\tilde{\lambda} \,.
 \label{gaugev}\err
 where $F_{ab}=2\,\der_{[a}V_{b]}$ is the field strength tensor.
 Notice that in this frame the fields $\tilde{f}$ and $\tilde{g}$ do not appear in the transformation rule
 $\delta_Q\,\tilde{\lambda}$.  Thus the corresponding Adinkra edge would be grey.
 Notice also that in this frame the field $\tilde{\chi}$ appears in the transformation rule
 $\delta_Q\,V_a$ only within a total derivative.
 Thus, if $V_a$ is interpreted as a gauge potential, subject to an equivalence under $V_a\to V_a+\der_a\,\alpha$,
 where $\alpha$ is a gauge parameter, then $\tilde{\chi}$ only contributes to $\delta_Q\,V_a$ as a gauge transformation.
 With these features in mind, we can represent the rules (\ref{gaugev}) using the following Adinkra,
 \brr \vC{\includegraphics[width=1.5in]{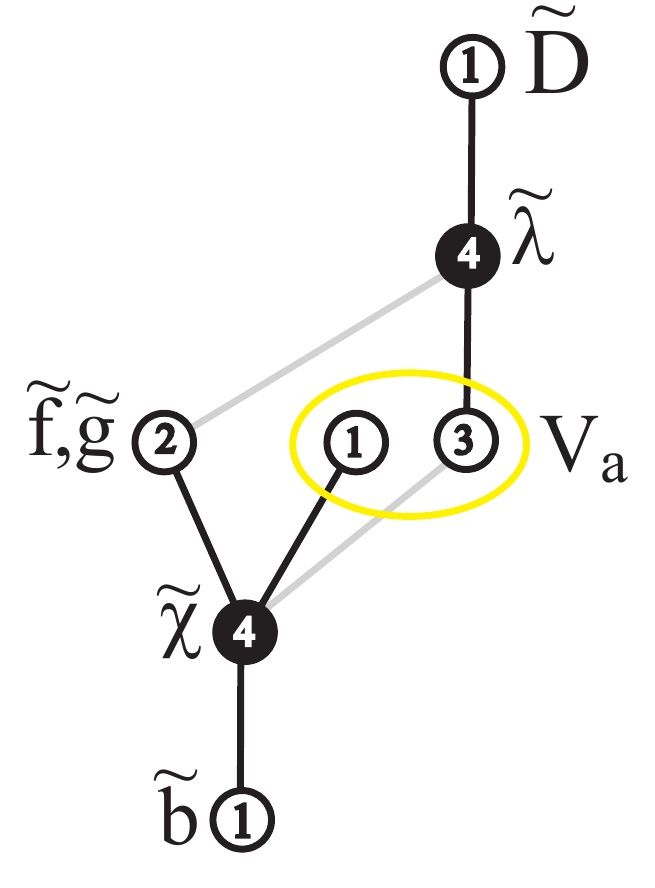}} \,,
 \label{vg}\err
 where we have represented the vector potential $V_a$ using two separate vertices: a singlet vertex
 codifying the gauge freedom, {\it i.e.} that part of $V_a$ which can be written as a total
 derivative, and another vertex codifying the gauge equivalence class.
 Since only the gauge part of $V_a$ ``talks back" to the lower half of the diagram (\ref{vg}), it is
 manifest on the diagram that the gauge-invariant field strength $F_{ab}:=2\,\der_{[a}V_{b]}$ resides in a submultiplet
 which does not involve any of the fields ``below" the grey lines.  In particular, the gauge invariant Adinkra
 can be read off of (\ref{vg}), and has the following form,
 \brr \vC{\includegraphics[width=1.5in]{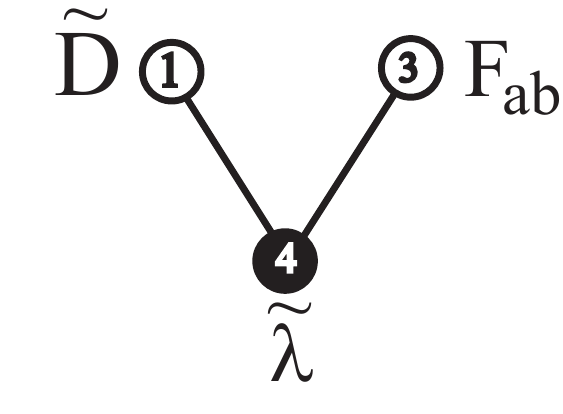}} \,.
 \label{gvv}\err
 The way this is done diagrammatically is by severing the grey edges in (\ref{vg}), indicating
 that the Wess-Zumino fields are set to zero, and then ``swiveling" the
 $V_a$ node upward two levels, pivoting on the $\lambda$ node, since such a maneuver codifies
 differentiation.  The $F_{ab}$ vertex describes three local degrees of freedom because
 this satisfies the Bianchi identity $\der_{[a}F_{bc]}=0$. \pagebreak
 The Gauge Vector transformation rules, which can be readily determined from (\ref{gaugev}),
 are
 \brr \d_Q\,\tilde{\lambda} &=&
      \fr12\,\gamma^{ab}\,F_{ab}\,\e
      -i\,\tilde{D}\,\gamma_5\e
      \nonumber\\[.1in]
      \d_Q\,\tilde{D} &=&
      \fr12\,\bar{\e}\,\gamma_5\dslash\,\tilde{\lambda}
      \nonumber\\[.1in]
      \delta_Q\,F_{ab} &=&
      -i\,\bar{\e}\,\gamma_{[a}\,\der_{b]}\tilde{\lambda} \,.
 \err
 It is these rules which are represented by the Adinkra (\ref{gvv}).

 Another way to reduce the Real Scalar multiplet is by imposing the constraint
 that the ``kite" fields in (\ref{vg}) each vanish.  This is done in two steps; first by writing
 $V_a=\tilde{V}_a+S_a$, where $S_a$ satisfies $\der_{[a}S_{b]}=0$ and is the part of $V_a$ which can
 be written as a total derivative, and then by imposing
 $\tilde{D}=0$, $\tilde{\lambda}=0$, and $\tilde{V}_a=0$.  This process ``switches off" the kite fields in
 (\ref{vg}), leaving behind the following gauge-invariant Adinkra
 \brr \vC{\includegraphics[width=1.5in]{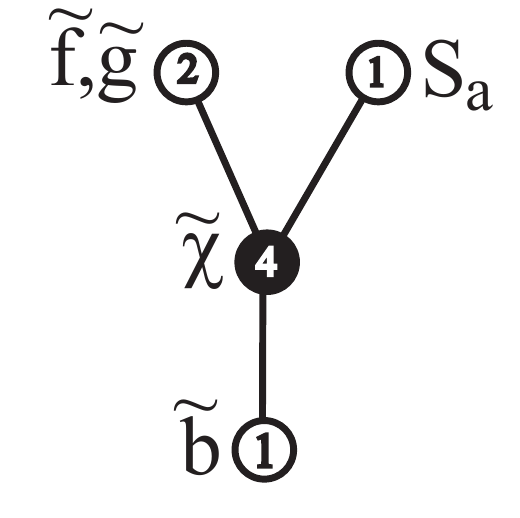}} \,,
 \err
 where the singlet vertex $S_a$ describes a closed one-form.
 This Adinkra describes the usual gauge superfield parameter in a supersymmetric
spin-1 gauge theory, and may be written in terms of superfields as
the sum of a Chiral multiplet and its Hermitian conjugate.

 Although the reduction of the Complex Scalar multiplet, equivalent to an unconstrained
 $N=1$ superfield, to its variety of
 submultiplets
 is very well known, both in terms of superspace and in terms of components, the perspective on this reduction
 described here is somewhat novel, we believe, both in terms of the judicious choices of frame redefinitions
 and in terms of how this matter plays out in terms of pictures.  Importantly, the discussion in this
 letter has allowed us to present the concept of grey Adinkra edges.  These indicate the appearance, in certain
 frames, of ``one-way" terms in supersymmetry transformation rules:
 ``upward-directed" terms which exist without the presence of ``downward-directed" counterparts.
 Such a feature played a role in our previous work in the context of $N=2$ supersymmetry \cite{RETM},
 and plays a role in related ongoing work, some of which which will appear in the near future.
 One purpose of this letter is to supply some independent elementary context and definitions to which
 we may refer in the future.  We also find the elucidation of the frame exhibited by (\ref{Chi_In})
 and (\ref{new_rules}),
 and its diagrammatic equivalent, shown in Figure \ref{k2}, adequately noteworthy. \\[.1in]

 \noindent{\bf{\Large Acknowledgements}}

  This research has been supported by the National Science Foundation Grant PHY-0354401,
  the endowment of the John S.~Toll Professorship, the University of Maryland Center for
String \& Particle Theory, National Science Foundation Grant
PHY-0354401, the University of Washington Royalty Research Fund,
and Department of Energy Grant DE-FG02-94ER-40854; M.F. is
grateful to the the Slovak Institute for Basic Research, Podvazie,
Slovakia, where much of this work was performed.

  \Refs{References}{[00]}
 \Bib{FG1} M.~Faux and S.~J.~Gates, Jr.:
 {\em Adinkras: A Graphical Technology for Supersymmetric Representation Theory},
  Phys.~Rev.~{\bf D71}~(2005),~065002;
  \Bib{Prepotentials}
  C.~Doran, M.~Faux, S.~J.~Gates, Jr., T.~H{\"u}bsch, K.~Iga, G.~Landweber:
  {\em Adinkras and the Dynamics of Superspace Prepotentials},
  Adv. S. Th. Phys., Vol. 2, no. 3 (2008) 113-164;
  \Bib{DFGHIL01} C.~Doran, M.~Faux, S.~J.~Gates, Jr., T.~H{\"u}bsch, K.~Iga, G.~Landweber:
 {\em On Graph Theoretic Identifications of
 Adinkras, Supersymmetry Representations and Superfields},
  Int. J. Mod. Phys. {\bf A}22 (2007) 869-930;
  \Bib{AT1}
  C.~Doran, M.~Faux, S.~J.~Gates, Jr., T.~H{\"u}bsch, K.~Iga,
  G.~Landweber, and R.~L.~Miller:
  {\em Topology types of Adinkras and the corresponding representations of $N$-extended
  supersymmetry},
  arXiv:0806.0050;
  \Bib{HDS}
   C.~Doran, M.~Faux, S.~J.~Gates, Jr., T.~H{\"u}bsch, K.~Iga,
  G.~Landweber, and R.~L.~Miller:
  {\em Relating doubly-even error-correcting codes, graphs, and
  irreducible representations of $N$-extended supersymmetry},
  arXiv:0806.0051;
 \Bib{RETM}
 C.~Doran, M.~Faux, S.~J.~Gates, Jr., T.~H{\"u}bsch, K.~Iga, G.~Landweber:
 {\it On the matter of $N=2$ matter},
 Phys. Lett. {\bf B}659 (2008) 441-446 \,;
 \Bib{Variant}
 S.J. Gates, Jr. and W. Siegel:
 {\em Variant Superfield Representations},
 Nucl.Phys. {\bf B}187 (1981) 389;
 \Bib{Nonminimal}
 B.~B.~Deo and S.J. Gates, Jr.:
 {\em Comments on nonminimal $N=1$ Scalar multiplets},
 Nucl.Phys. {\bf B}254 (1985) 187-200;
 \Bib{Superspace}
 S.~J.~Gates, M.~T.~Grisaru, M.~Rocek and W.~Siegel,
 {\it Superspace, or one thousand and one lessons in supersymmetry},
  Front.\ Phys.\  {\bf 58}, 1 (1983)
 \Bib{Wess_Bagger}
 J.~Wess and J.~Bagger,
 {\it Supersymmetry and supergravity},
  Princeton Univ. Pr. (1992) \,;
 \endRefs

 \end{document}